# Manipulating Kondo Temperature via Single Molecule Switching

*Violeta Iancu, Aparna Deshpande, Saw -Wai Hla\**
*Nanoscale & Quantum Phenomena Institute, Physics & Astronomy Department,*
*Ohio University, Athens, OH 45701, USA.*

\*Corresponding author: Email: hla@ohio.edu,
Web: www.phy.ohiou.edu/~hla

**Two conformations of isolated single TBrPP-Co molecules on a Cu(111) surface are switched by applying +2.2 V voltage pulses from a scanning tunneling microscope tip at 4.6 K. The TBrPP-Co has a spin-active cobalt atom caged at its center and the interaction between the spin of this cobalt atom and free electrons from the Cu(111) substrate can cause a Kondo resonance. Tunneling spectroscopy data reveal that switching from the saddle to a planar molecular conformation enhances spin-electron coupling, which increases the associated Kondo temperature from 130 K to 170 K. This result demonstrates that the Kondo temperature can be manipulated just by changing molecular conformation without altering chemical composition of the molecule.**

KEYWORDS: *Nanoscience and Technology, Scanning Tunneling Microscopy, Molecule Manipulation, Kondo, Molecule Switch.*

---

The interaction between a magnetic impurity and electrons from a nonmagnetic metal environment gives rise to a resonance near the Fermi level, known as the Kondo effect[1-3]. With the technological advances in nanoscale fabrication the Kondo effect has seen a revival of interest in the recent years[4-10] and has been observed in a wide variety of systems, ranging from semiconductor quantum dots[9], single atoms[2-4] to carbon nanotubes[10]. Among them, molecules exhibiting magnetic properties are of special interest to investigate the spin-electron interaction because of their potential in spintronic applications[11]. Here, we perform a comparative study of isolated molecules of TBrPP-Co and TBrPP-Cu [5, 10, 15, 20 – Tetrakis -(4-bromophenyl)-porphyrin-M (M = Co or Cu)] on a Cu(111) surface using low temperature scanning tunneling microscopy and spectroscopy. The TBrPP-Co and TBrPP-Cu molecules are composed of a porphyrin unit with a cobalt (Co) or copper (Cu) atom caged at the center and four bromophenyl groups at the end parts.

The experiments were performed by using a home-built ultrahigh vacuum (UHV) low-temperature STM system operated at 4.6 K[12]. The Cu (111) sample was cleaned by repeated cycles of Ne ion sputtering and annealing to 700K. An electrochemically etched polycrystalline tungsten wire was used as the STM-tip. The tip apex was coated with copper in-situ by making a tip-sample mechanical contact on a bare Cu(111) terrace prior to the spectroscopic measurements[13]. For a comparative study, TBrPP-Co and TBrPP-Cu [5, 10, 15, 20 – Tetrakis -(4-bromophenyl)-porphyrin-Co or Cu] molecules were separately deposited on a clean Cu(111) surface held at ~100 K by thermal evaporation using a homebuilt Knudsen cell. Then the sample was transferred to the STM chamber without breaking the UHV conditions and the sample temperature was lowered to 4.6 K for the experiment.

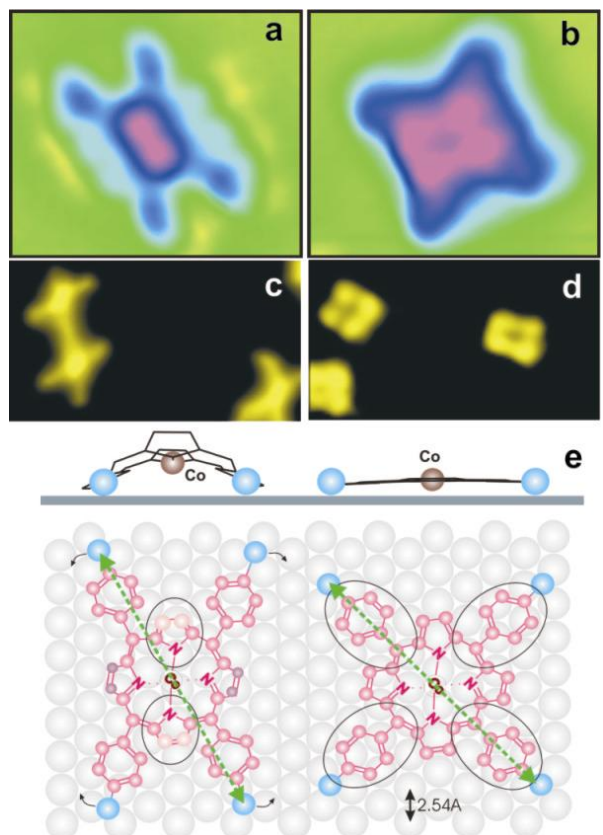

**Figure 1.** Molecular conformations. (**a**) STM images of saddle conformation (width ~11Å, length ~18 Å) and (**b**) planar conformation (length ~ 15.5 Å) of TBrPP-Co on Cu(111). Similar conformations are observed for the TBrPP-Cu on Cu(111); the saddle (**c**), and planar (**d**), conformations. (**e**) The corresponding models of saddle (left) and planar (right) conformation. Blue and pink color balls represent bromine and carbon atoms, respectively. The encircled regions of the molecule indicate the area providing higher current in the STM images. The length of the molecule along the diagonal (indicated with a dashed green arrow) is unchanged upon switching the two conformations. [Imaging parameters: $V_t$ = 1 V, $I_t$ = 250 pA]

TBrPP-Co molecules anchor on the Cu(111) surface via their four bromine atoms positioning at the three-fold hollow sites of the copper surface and form two molecular conformations: the saddle and the planar (Fig. 1). Metallo-tetraphenyl-porphyrin molecules are known to have several conformations including the saddle and planar both in gas phase and in solutions[14,15]. On the Cu(111) substrate 25% of the molecules are found adsorbed in the saddle conformation. In the saddle conformation, the central part of molecule is bent by the elevation of two pyrrole units of the porphyrin macrocycle. The STM images acquired at 4.6 K substrate temperature under an ultra-high-vacuum environment show two protrusions for the saddle (Fig. 1a), presumably originated from the two lifted pyrrole units. The planar molecule has an approximately square shape and the porphyrin plane is positioned parallel to the Cu(111) surface. In this position, the four bromophenyl groups of the molecule can interact strongly with the surface via π-interactions. The STM images of planar TBrPP-Co reveal a set of four lobes (Fig. 1b).

The distortion (saddling) of the molecule can be removed by applying a burst of voltage pulses from the STM-tip, which supplies the necessary energy to switch the molecule from the saddle to the planar conformation. To switch the molecule (Fig. 2), the STM-tip is positioned at a fixed height above the center of a saddled molecule, and a fixed voltage pulse of +2.2 V is applied for a few seconds. The switching event is directly observed by monitoring the changes in tunneling current. The subsequent STM images confirm that the molecule switches into a planar conformation. The tunneling current signals during the switching event mostly include two discrete steps (Fig. 2a) indicating that a two step-switching mechanism has been involved. By varying the time duration of the applied voltage pulses these two steps can be individually induced. Now, we observe switching of molecule into an intermediate step with a corresponding single-step current signal. The molecule can then be conformed into the planar by providing another burst of voltage pulses. Figure 2 captures the switching steps: First, half of the porphyrin unit is switched from saddle to planar with two pronounced lobes on one side while the other side of the molecule remains in the saddle position. The supply of another voltage pulse completes the task by flattening the rest of the porphyrin unit, which now appears as a four-lobe (planar) conformation. The molecule remains intact after switching. This is verified by laterally moving the molecule with the STM tip across the surface and by dissociating the bromine atoms from the four bromophenyl groups using higher voltages[12]. A reverse switching[16] from planar to saddle conformation requires voltage pulses higher than 3 V for this molecule. Thus, it is difficult to achieve this in the current experimental framework as higher energy transfer can lead to the fragmentation of the molecules.

For the study of Kondo effect on conducting surfaces, a scanning tunneling microscope (STM) is an ideal tool to detect the localized spin-electron interaction at an atomic limit by measuring the energy dependence of the local density of states (LDOS) around the Fermi level[2,3]. The shape and width of the resonance gives insight into the tunneling process through the impurity and determines the Kondo temperature of the system[17-19]. We investigate the spin-electron interaction between isolated TBrPP-Co molecules and the free electrons of Cu(111) by measuring differential conductance (dI/dV) tunneling spectroscopy over both molecular conformations. During this process, the STM tip is positioned at the center of the molecule and the voltage is ramped between ±60 mV. The resultant dI/dV spectra of both saddle and planar molecules consistently reveal a small 'dip' located around the substrate Fermi level, i.e. zero tunneling bias, which we attribute as the Kondo resonance (Fig. 3). The dI/dV measurements were repeated with different tips on a large number of molecules for consistency. We measure the Kondo signatures before and after the switching events, as well as over the naturally adsorbed molecules. Fig. 3a presents a representative Kondo signature of the saddle TBrPP-Co. The experimental data is fitted by using a model derived by Ujsaghy et al [18]. The resultant Kondo temperature of saddle TBrPP-Co at 4.6 K is 130 ± 15 K. The width of Kondo resonance increases for the planar TBrPP-Co and the corresponding Kondo temperature is determined as 170 ± 10 K. The errors stated here represent the statistical deviation of the measured data.

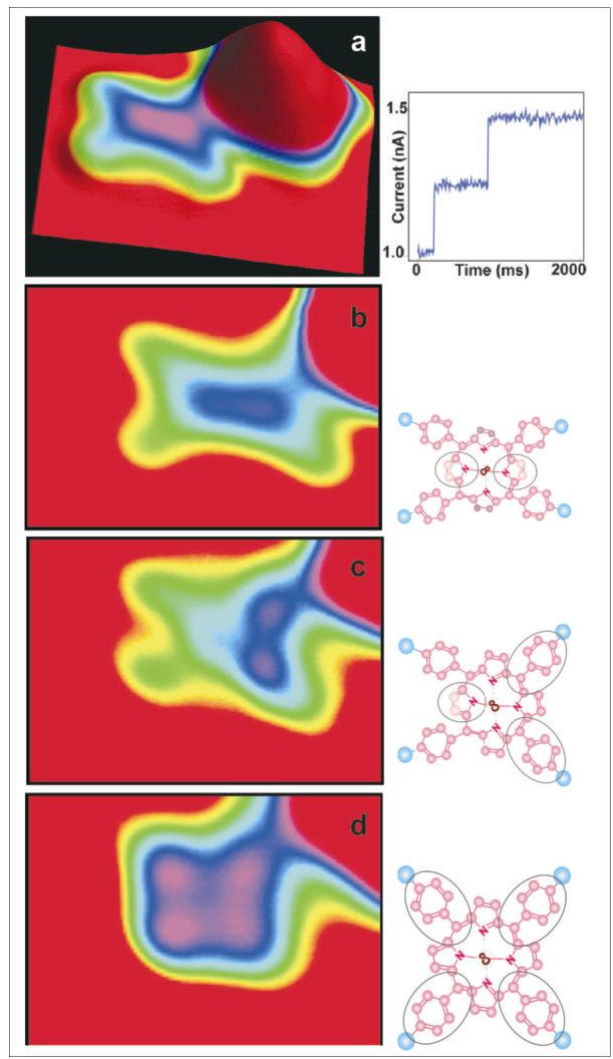

**Figure 2.** Single molecule switch. **(a)** A three dimensional STM image shows a TBrPP-Co molecule next to a Cu cluster used here as a landmark. Half of the porphyrin unit from **(b)** is switched into the planar conformation **(c)** by applying a voltage pulse. **(d)** Another voltage pulse changes the molecule into a four lobe (planar) conformation. [Imaging parameters: $V_t$ = 1 V, $I_t$ = 840 pA]

To further verify the origin of the observed resonance around Fermi energy, we comparatively investigate the dI/dV spectra of isolated TBrPP-Cu molecules on Cu(111). The TBrPP-Cu has a similar molecular structure as the TBrPP-Co except that the Co atom is replaced by a nonmagnetic copper (Cu) atom. Since the copper is nonmagnetic, the Kondo effect caused by spin-electron interaction should not occur. The TBrPP-Cu molecules adsorb on Cu(111) with two different conformations as in the case of TBrPP-Co (Fig 1c and 1d). As expected, the dI/dV spectra of both saddle and planar conformations of isolated TBrPP-Cu molecules are featureless around the substrate Fermi level (Fig. 3c and 3d). This verifies that the observed resonances in both conformations of TBrPP-Co originate from the spin-electron interactions, i.e. the Kondo effect.

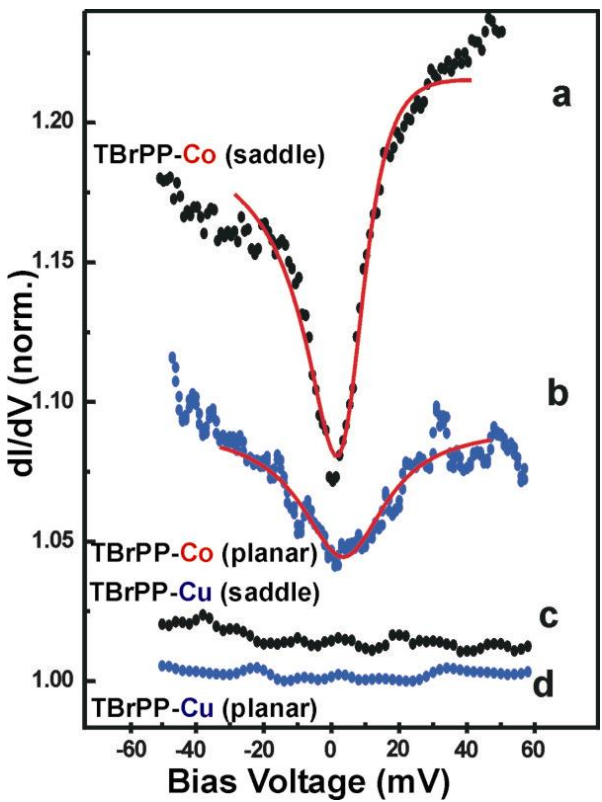

**Figure 3.** Kondo signatures. **(a)** A conductance spectrum of the saddle TBrPP-Co molecules showing a dip (Kondo resonance) around surface Fermi level (0V). The solid line represents the Fano line-shape fit to the data. The fit parameters for this curve are q = 0.29 ± 0.01, $\varepsilon_K$ = 4.52 ± 0.29 mV. **(b)** The width of Kondo resonance increases in the dI/dV spectra of planar TBrPP-Co molecules, which corresponds to an increase in Kondo temperature. The solid line represents the Fano line-shape fit to the data. The fit parameters are q = 0.01 ± 0.03, $\varepsilon_K$ = 3.70 ± 0.97 mV. **(c), (d)** The dI/dV spectra of saddle and planar TBrPP-Cu molecules do not reveal any features around the surface Fermi level, i.e. 0V. dI/dV spectra are taken with a lock-in amplifier with an AC modulation of 4mV r.m.s at 500Hz. The spectra for TBrPP-Co saddle, TBrPP-Co planar and TBrPP-Cu saddle have been shifted vertically by 0.07, 0.04, and 0.01, respectively. All the spectra are taken at the center of each molecule.

In case of adsorbed single Co atoms on Cu(111) surface, the Kondo temperature is measured as ~53 K[4]. The observed high Kondo temperature in TBrPP-Co molecules is caused by enhanced coupling of the magnetic impurity, caged by the porphyrin molecule, to the conduction electrons of the surface through the molecular bonding[5,6]. To explain the increased Kondo temperature after switching, we determine the electronic structures of both molecular conformations by measuring the orbital-mediated tunneling spectra (OMTS) over isolated molecules (Fig. 4)[20,21]. All the spectra are measured at the center of the molecules. Similar OMTS spectra of TBrPP-Cu molecules are also taken for a comparative study. The OMTS of saddle TBrPP-Co (Fig. 4a) reveals a rather large gap between the highest occupied and lowest unoccupied molecular orbitals (HOMO-LUMO). The HOMO of saddle TBrPP-Co on Cu(111) is located at -1.55 eV below the Fermi level (6.26 eV from the vacuum level). At the positive bias, only the onset of the lowest occupied molecular orbital (LUMO) can be observed in this voltage range. On the contrary, the OMTS of planar TBrPP-Co (Fig. 4b) shows a HOMO-LUMO gap of 1.46 eV, which is comparable to that of electrochemical measurement[22], 1.4 eV, on planar cobalt porphyrin (TPP-Co). The first occupied orbital of planar TBrPP-Co is located at 0.7 eV below the Fermi level (5.59 eV relative to the vacuum level). The origin of this orbital is assigned to the ionization of $d_z^2$ orbital of the Co atom, which is absent in case of planar TBrPP-Cu (Fig. 4d). The interesting case is that this orbital is also absent in the OMTS spectra of saddle TBrPP-Co (Fig. 4a). In the saddle conformation the porphyrin plane is bent and the Co atom is lifted away from the surface. This effectively reduces a direct spin-electron coupling between the Co atom and the surface state free electrons of Cu(111) resulting in the disappearance of the Co $d_z^2$ associated peak from the spectrum of the saddle TBrPP-Co. As a result, the Kondo temperature is decreased in case of saddle as compared to planar TBrPP-Co. Both molecule conformations bind to the substrate via bromine atoms and hence, spin-electron coupling through molecular bonding is still permitted in the saddle conformation, which provides the observed Kondo temperature of 130 K. Our measurements indicate that the molecular Kondo effect may be generated by more than a single path, if we consider the spin-electron coupling through molecular bonding as one path and a direct coupling of Co $d_z^2$ orbital to the substrate as the other path. In case of saddle conformation, the Kondo effect occurs most likely through a single path, coupling via molecular bonding, while the higher temperature Kondo effect in the planar case involves both paths.

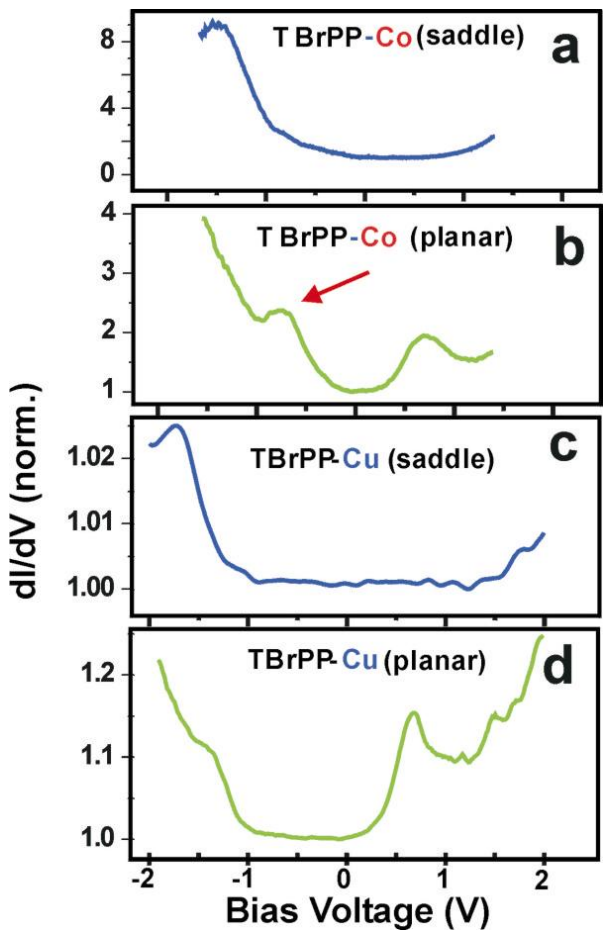

**Figure 4.** The orbital mediated tunneling spectra (OMTS). **(a)** The STM-OMTS spectra of saddle (blue) and **(b)** planar (green) TBrPP-Co. The highest occupied molecular orbital (indicated with an arrow) in the planar molecule data is originated from the Co atom, which is absent in the saddle spectrum (blue). **(c), (d)** The STM-OMTS spectra of saddle and planar conformations of TBrPP-Cu.

In summary, we show that the two different Kondo temperatures can be changed via a single molecule switching mechanism. The increase in Kondo temperature after switching is explained from the OMTS spectra, which reveal a stronger Co-surface interaction in case of the planar conformation. Our finding of two different Kondo temperatures, which can be tuned just by switching the conformations without destroying or altering molecule's chemical compositions may have a distinct advantage in development of novel spintronic devices in the future.

**Acknowledgement:** We thank A.O. Govorov and S.E. Ulloa for useful discussions. We gratefully acknowledge the financial supports provided by the United States Department of Energy grant number DE-FG02-02ER46012, and National Science Foundation NIRT grant number DMR-0304314.